\documentclass[aps,prl,reprint,amsmath,amssymb]{revtex4-1}

\usepackage{amssymb,amsfonts,amsmath,graphicx}
\usepackage{color}
\usepackage{dsfont}

\def\be{\begin{equation}}
\def\ee{\end{equation}}
\def\bea{\begin{eqnarray}}
\def\eea{\end{eqnarray}}

\def\p{\partial} 
\def\nn{\nonumber}
\def\f{\frac}

\def\l{\left(}
\def\r{\right)}

\def\grad{\vec{\nabla}}

\newcommand{\overbar}[1]{\mkern 1.5mu\overline{\mkern-1.5mu#1\mkern-1.5mu}\mkern 1.5mu}

\newcommand{\Lagr}{\mathcal{L}}

\usepackage[
margin=0.6in,
]{geometry}
\begin{document}

\title{
Emergent Structures in an Active Polar Fluid : dynamics of shape, scattering and merger}

\author{Kabir Husain}
\affiliation{%
Simons Centre for the Study of Living Machines, National Centre for Biological Sciences (TIFR), Bellary Road, Bangalore 560 065, India
}%
\author{Madan Rao}
\email{madan@ncbs.res.in}
\altaffiliation{%
on lien from {\it Raman Research Institute, C.V. Raman Avenue, Bangalore 560080, India}
}%
\affiliation{%
Simons Centre for the Study of Living Machines, National Centre for Biological Sciences (TIFR), Bellary Road, Bangalore 560 065, India
}%

\date{\today}

\begin{abstract}
Spatially localised defect structures emerge spontaneously in a hydrodynamic description of an active polar fluid comprising polar `actin' filaments and `myosin' motor proteins that (un)bind to filaments and exert active contractile stresses. These emergent defect structures are characterized by distinct textures and can be either static or mobile -  we derive effective equations of motion for these  `extended particles' and analyse their shape, kinetics, interactions and scattering. Depending on the impact parameter and propulsion speed, these active defects undergo elastic scattering or merger. Our results are relevant for the dynamics of actomyosin-dense structures at the cell cortex, reconstituted actomyosin complexes and 2d active colloidal gels. 
\end{abstract}

\maketitle

\indent The dynamical interplay between topological defects, elasticity and flow has been a recurrent theme in the physics of condensed media \cite{chaikinlubensky}. This has been of recent interest in the context of active systems, in which activity drives the formation and dynamics of collective or emergent structures: be it the rotation of spiral defects in polar gels \cite{ActiveReview,kruse}, the translation, ordering and complex flows of $+1/2$ disclinations in active nematics composed of filament-motor complexes \cite{sanchez2012spontaneous,keber2014topology,decampOrientationOrderDefects,eliasdefectordering,GiomiExcitablePatterns,pismendefectdynamics,giomidefectdynamics,giomiAnalyticDefects,yeomansFluxLattice} or the spontaneous formation of interacting mobile aggregates of active colloidal particles \cite{ChaikinColloidalSurfers, BocquetDynamicClustering, lowenDynamicClustering,LowenClusteringCollapse,LowenActiveCrystals}. These structures are often particle-like, with a well-defined position and orientation, and it is desirable to cast their dynamics solely in terms of their own coordinates and form factor

\indent In this manuscript we analyse the emergent dynamics of polar filaments (`actin') actively driven by motor proteins (`myosin') that undergo turnover, i.e., binding/unbinding (with an estimated timescale of $\sim 6$s \textit{in vivo} \cite{turnoverCharras}) \cite{LeeKardar,SumithraMenon,AhmadiCoarseGrain,Gowrishankar}. Recent studies on actomyosin-dependent molecular clustering at the cell surface and in in-vitro reconstitutions of a thin actomyosin layer on a supported membrane, provide our primary motivation \cite{Cell2008,Ambika,Gowrishankar,AmitFDPO}. Our main results are: (i) A variety of spatially compact structures, such as mobile virtual defects, rotating spiral defects or stationary asters, emerge spontaneously from an interplay between elastic, dissipative and active stresses. (ii) Activity drives defect motion and shape deformation, with isolated virtual defects moving as {\it extended particles}, dressed by a cloud of myosin density. (iii) These extended particles interact via the accompanying myosin cloud, and exhibit both {\it phoresis} and {\it taxis}. (iv) We study the dynamics, elastic scattering and (inelastic) merger of a pair of mobile defects as a function of impact parameter and self-propulsion velocity, by casting the effective dynamics in terms of defect coordinates and shape factor. We note that all the emergent characteristics of the extended particle - its integrity, dynamics and their mutual interactions, are consequences of the dressed myosin cloud.

\indent Our results are obtained by solving hydrodynamic equations for a thin film of active fluid comprising actin filaments, described by a concentration $c$ and orientation $\mathbf{n}$, and filament-bound myosin motors, with concentration $\rho$, which dissipate momentum via frictional damping at  the cell substrate. As found from systematic coarse-grainings \cite{AhmadiCoarseGrain}, the active contributions to the dynamics of $c$ and $\mathbf{n}$ come about {\it only in the presence of bound myosin} - the Supplementary contains a phenomenological derivation of the following simplified hydrodynamic equations, based on symmetry and conservation laws \cite{ActiveReview,TonerTu},
\bea
\label{eq:ceqn}
&& \f{\p c}{\p t} =  -\grad\cdot \l -D_a\grad c  +  v_0 c \mathbf{n} - W c \grad \rho \r  \\
\label{eq:rhoeqn}
&& \f{\p \rho}{\p t}   =  -\grad\cdot \l -D_m\grad \rho + v_0\rho \mathbf{n} \r +  k_b\f{c}{c+c_h} - k_u \rho  \\
&&\frac{\p \mathbf{n}}{\partial t} + \lambda(\rho) (\mathbf{n}\cdot\grad) \mathbf{n}  =  K \nabla^2\mathbf{n} 
- \zeta_0\grad c + \zeta\grad\rho
 \nn \\
&&  + (\alpha(c) - \beta(c) n^2)\mathbf{n}
\label{eq:neqn}
\eea
where we have taken the active stress  $\mathbf{\sigma}^{act}= -W(c) \rho$ ($W<0$, for contractility), and we work in the one-constant approximation \cite{degennes}. We take $\alpha(c) = \nu(c/c_{\ast} - 1)$ and $\beta(c) = \nu(1+c/c_{\ast})$, which ensures that $\vert {\bf n}\vert^2 \to 1$ when $c \gg c_{\ast}$ \cite{aparnadynamicself}. In Eq.\,\ref{eq:neqn}, $\zeta_0 > 0$ ensures positive filament compressibility, while $\zeta > 0$ describes the preferential alignment of the filament orientation to gradients of myosin density \cite{SumithraMenon}, here a consequence of the active stress. These equations generalise earlier models of acto-myosin \cite{LeeKardar,SumithraMenon,Gowrishankar,aparnadynamicself,AhmadiCoarseGrain}, which had been proposed on the strength of both symmetry arguments and coarse grainings, by explictly treating both the polar filaments and the myosin motors as dynamical variables. We note a resemblence to recently proposed models of chemoactive colloids \cite{catesColloids}, which suggests that our results may be of wider applicability.

\begin{figure}
\begin{centering}
\includegraphics[scale=.04]{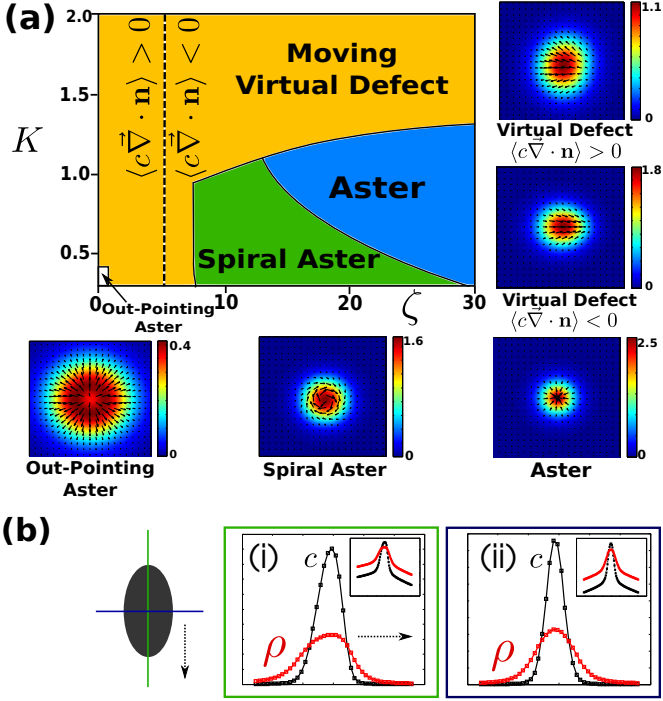}
\end{centering}
\caption{\label{fig:textures} \textbf{(a)} Typology of spatially compact single-defect structures in the $K$ - $\zeta$ plane for a fixed concentration of filaments $c_0$, together with snapshots of the textures formed by the filament orientation $\mathbf{n}$ - the heat map represents $c$ and the arrows represent $c\mathbf{n}$. All snapshots are equally sized. \textbf{(b)} Myosin (red) and actin (black) concentration profiles within a moving virtual aster, along directions (i) longitudinal and (ii) 
transverse to the direction of motion (dashed arrow). Insets: semi-log scale.
}
\end{figure}

\indent We begin by studying the linear instability of the homogeneous, orientationally isotropic phase when $c_0 < c_{\ast}$, under the perturbations, $c(\mathbf{r},t) = c_0 + \delta c(\mathbf{r},t)$, $\rho(\mathbf{r},t) = \rho_0 + \delta \rho(\mathbf{r},t)$ and $\mathbf{n}(\mathbf{r},t) = \delta \mathbf{n}(\mathbf{r},t)$, where $\rho_0 = \f{c_0}{c_0 + c_h}k_b/k_u$. We find two potential instability mechanisms : (i) the binding of myosin and the contractile advection of filaments ($\propto W c \grad \rho$) induces an density clumping instability when $k_b |W| \gtrsim D_ak_u$, with a band of unstable wavevectors between $k = 0$ and $k = \sqrt{|W|k_bc_0 - D_a k_u}/D_a$. Filament orientation remains disordered ($\langle \mathbf{n}\rangle= 0$). (ii) Myosin-induced torques ($\propto \zeta \grad\rho$) and advection ($\propto v_0 c \mathbf{n}$) drives a splay deformation ($|\grad\cdot\mathbf{n}| > 0$) and density clumping when $D_a |\alpha(c_0)| \lesssim v_0\zeta\rho_0$, with a band of unstable wavevectors between $k = 0$ and $k = \sqrt{\f{D_a k_u \alpha(c_0) + c_0 k_b v_0 \zeta}{D_a K k_u}}$. In both cases the instability towards clumping is non-oscillatory \cite{suppinfo}.

\indent To see the fine structure of the emergent compact configurations, we need to go beyond the linear regime. We numerically integrate Eqs.\,\ref{eq:ceqn}-\ref{eq:neqn} on a square grid with periodic boundary conditions using an explicit Forward-Time-Central-Space (FTCS) scheme and appropriate initial conditions \cite{suppinfo}. Starting from a homogeneous, isotropic phase, we observe that the system quickly settles down into an evolving population of spatially compact actin-myosin clusters (Movies\,1,\,2 \cite{suppinfo}). To explore the typology of compact structures formed, we choose initial conditions that produce a single defect and build a phase diagram in $K$ and $\zeta$.

\indent Figure\,\ref{fig:textures}(a) shows the single-defect phase diagram together with the compact density profiles and orientational textures that we obtain. All structures consist of a compact accumulation of actin filaments dressed by a wider myosin cloud (Fig.\,\ref{fig:textures}(b)). At low $K$ they are radially symmetric with zero net polarity ($\langle c \mathbf{n} \rangle = 0$) and hence static. The texture is set by the control parameter $\zeta/\zeta_0$. When $\zeta/\zeta_0$ is small (large) the term $\zeta_0 \grad c$ ($\zeta \grad\rho$) dominates to form an out(in)-pointing aster. As we show in the SI, integrating out the myosin density when it is `fast', in the limit of large $\zeta/\zeta_0$, results in effective equations for the active filament identical to those analyzed in \cite{Gowrishankar,KripaPhase}, with a `negative compressibility' in the $\mathbf{n}$-equation \cite{aparnadynamicself,KripaPhase}.

\begin{figure*}
\begin{centering}
\includegraphics[scale=.14]{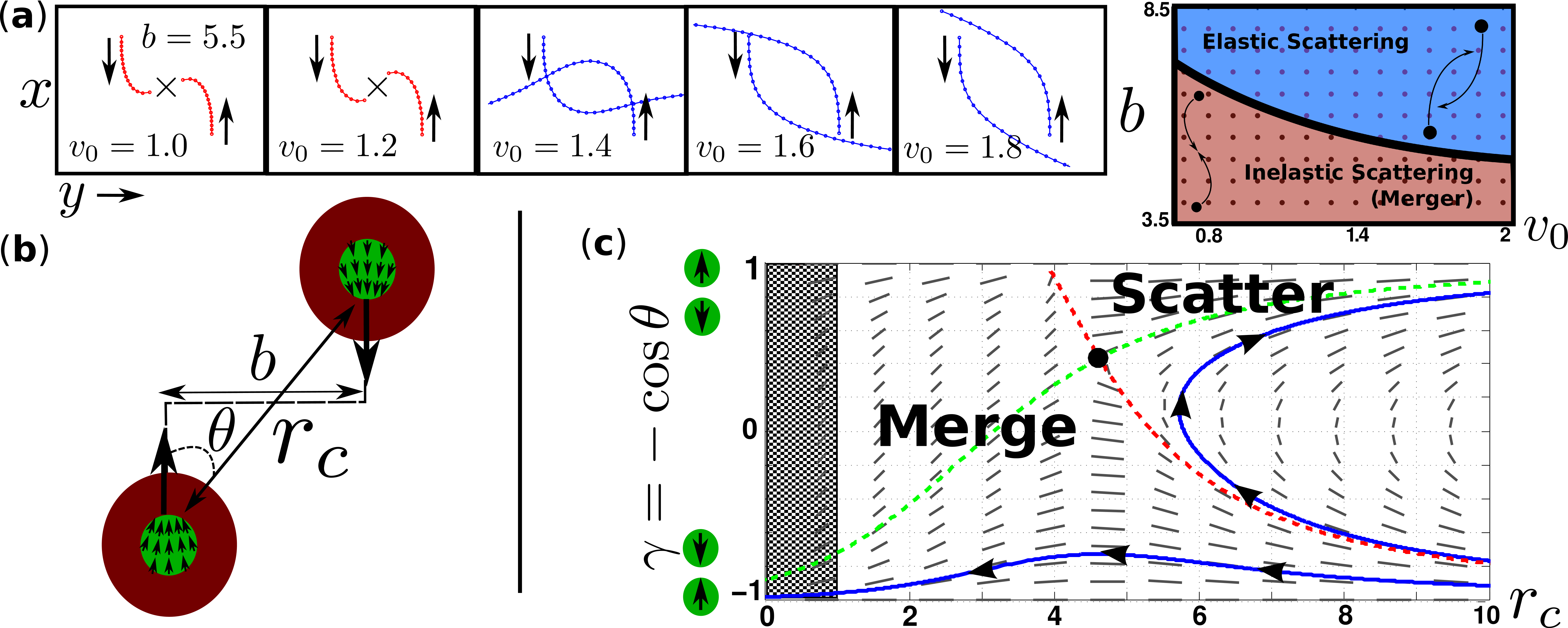}
\end{centering}
\caption{\label{fig:scatter} \textbf{(a)} Left - several representative trajectories from the numerics with varying $v_0$ and fixed $b = 5.5$ (merger is depicted as a $\times$). Right - The outcome as a function of impact parameter $b$ and self-advection $v_0$. \textbf{(b)} Left - The scattering geometry (green: $c$, arrows: $\mathbf{n}$, red: $\rho$). \textbf{(c)} The phase portrait of the scattering dynamical system, Eqs.\,\ref{eq:dynamicalsystem}, generated using the software \textit{pplane} (http://math.rice.edu/~dfield/dfpp.html). The black dot is the saddle point whose unstable separatrix (red dashed line) separates the merge region from the scattering region. Plotted in green is the stable separatrix. In blue are two example trajectories. The shaded region is the distance at which the merger will occur.}
\end{figure*}

At an intermediate value of $\zeta/\zeta_0$, $\zeta_0\grad c$ dominates at the core of the domain and $\zeta \grad \rho$ dominates at the boundary, forming a spiral aster texture, which rotates with an angular velocity $\omega(r) \propto v_0 n_{\theta}(r)$ \cite{ActiveReview,kruse}.

\indent When $K$ is large the defect core moves abruptly from within to outside the compact domain,
giving rise to an \textit{apparent} or \textit{virtual defect} \cite{PetteyLubensky,SethnaTextures,SarasijPragyaTexture}. Depending on the relative strength of $\zeta\grad\rho$ and $\zeta_0\grad c$, the virtual defect is either a virtual out-pointing-aster ($\langle c \grad\cdot\mathbf{n} \rangle > 0$) or a virtual aster ($\langle c \grad\cdot\mathbf{n} \rangle < 0$).

\indent These textures can be qualitatively understood as arising from a competition between the first three terms on the R.H.S. of Eq.\,\ref{eq:neqn}. We further note that the dynamics of $\mathbf{n}$ (when $\lambda = 0$) can be obtained as $\p_t \mathbf{n} = -\delta \mathcal{L}/\delta \mathbf{n} $, where the Lyapunov function $\mathcal{L} = \int d^2x \left[\mathop{} K \l \p_i n_j \r \l \p_i n_j \r   + (\zeta \rho - \zeta_0 c)\grad\cdot\mathbf{n}\right]$. The quantity $\zeta \rho - \zeta_0 c$ thus promotes a spontaneous splay \cite{SarasijPragyaTexture,PragyaCylinder}, with a negative (positive) divergence when $\zeta \rho - \zeta_0 c > 0$ ($<0$). The Lyapunov function $\Lagr$ is analogous to the free energy of textured Langmuir domains \cite{SarasijPragyaTexture,PetteyLubensky}, indicating that the transition between aster and virtual defect is first order, as a function of the control parameter $R_a(\zeta \rho_0 - \zeta_0 c_0)/K$ \cite{PetteyLubensky}, where $R_a$ is the radius of the filament domain.

\indent The profile of filament and myosin densities is set by the interplay between the binding and advective nonlinearities and it is difficult to obtain an exact expression. However, we may exploit the compactness of the filament domain to analyse the decay of the dressed cloud of myosin. Compactness implies that $c(\mathbf{r})$, $|\mathbf{n}| \approx 0$ for $r > R_a$ (where $R_a$ is naturally interpreted as the radius of the filament domain). The outer solution (i.e., for $r > R_a$) for $\rho$ at steady state satisfies: $D_m \nabla^2 \rho = k_u \rho$. The radially symmetric solution is found to be,
\be \label{eq:profile}
\rho(r) = A_1 \text{K}_0\l r/r_0 \r
\ee
where $\text{K}_{\nu}(r)$ is the modified Bessel function of the second kind and $r_0 = \sqrt{D_m/k_u}$. The constant $A_1$ may be determined by matching the inner solution ($r<R_a$, where $|\mathbf{n}|$, $c \geq 0$) with the outer solution.

\indent The sudden appearance of a net polarity ($|\langle c \mathbf{n} \rangle| > 0$) manifests as a shape distortion away from circularity and a centre-of-mass motion driven by the advective current, $\propto c \mathbf{n}$, which we analyse to lowest order in deformation. We transform to a comoving frame with velocity $\mathbf{v_c}$ and solve for the lowest order axisymmetric steady-state solution: i.e. of the form $\rho_0(r) + \rho_1(r) \cos \theta$, where $\theta = 0$ is the direction of motion. When deviations from circularity are small ($\rho_1 \ll \rho_0$), $\rho_0$ is given by Eq.\,\ref{eq:profile} and
\be \label{eq:axiprofile}
\rho_1(r) = A_2 \text{K}_1 \l \f{r}{r_0} \r - \f{v_c}{\sqrt{D_mk_u}} A_1 \, r \, \text{K}_0\l \f{r}{r_0} \r 
\ee
where $A_2$ may once again be obtained from matching the inner and outer solutions at $r = R_a$.

\indent From our numerical analysis, we find that an isolated moving virtual defect with its myosin cloud maintains its shape and texture over time. The density and orientation fields then obey a traveling front form $c=c({\bf r}-{\bf R}(t))$, $\mathbf{n}=\mathbf{n}({\bf r}-{\bf R}(t))$, $\rho=\rho({\bf r}-{\bf R}(t))$. We may thus describe the motion of a single virtual defect, isolated from others, solely in terms of its centre-of-mass coordinate $\mathbf{R}(t)\equiv c_{t}^{-1}\displaystyle\int_{\Omega} d^2r \mathop{} \mathbf{r} c$ and net polarity $\mathbf{p}(t)\equiv c_{t}^{-1}\displaystyle\int_{\Omega} d^2r \mathop{} \mathbf{n} c$, where $c_{t} = \displaystyle\int_{\Omega} d^2r \mathop{} c$ and $\Omega$ is the compact support of the defect,
\bea \label{eq:eff_general}
&& \mathbf{\dot R}(t) = \f{1}{c_t}\int d^2r \mathop{} \mathbf{J_f} \approx \f{v_0}{c_t}\int d^2r \mathop{} c \mathbf{n} = v_c \hat{\mathbf{p}}(t) \nn \\
&& \mathbf{\dot p}(t) = \f{1}{c_t}\int d^2r \mathop{}c \underbrace{\l \f{\mathbf{J_f}}{c}\cdot\grad + \f{\p}{\p t} \r}_{\text{convective derivative}}\mathbf{n} = 0
\eea
where the overdot is a time derivative. Above, we have made use of the fact that filament diffusion $D_a \grad c$ and the $W c \grad\rho$ term are resposible for maintaining the size of the moving domain, and hence they balance to $0$ everywhere. This leaves only the advective contribution ($\propto v_0 c \mathbf{n}$) in the first equation.

\indent The speed of the domain, $v_c = v_0 |\mathbf{n}^c|$, may be approximately calculated from an \textit{ansatz} (shown in Fig.\,S3(b) in \cite{suppinfo}) and is found to increase from $\sim 0.85v_0$ to $v_0$ as the apparent defect core moves further from the centre of the domain. This is consistent with the numerics \cite{suppinfo}, though we note that the asymmetric profile of the filament and myosin fields will also contribute to the speed of the domain.

\indent We next investigate the interaction between these `particles'. When the filament cores of two domains are well-separated, $|\mathbf{n}| \approx 0$ in the region between them and elastic interactions may be neglected. Their interaction is instead mediated by the associated myosin clouds  - stationary defects do not interact unless they are close enough to allow for a finite overlap between their myosin clouds, with a interaction length scale $r_0 = \sqrt{D_m/k_u}$ - this interaction leads to their merger (Movie\,1 \cite{suppinfo}). On the other hand, moving virtual defects may initially be well separated and then approach each other (Movie\,2 \cite{suppinfo}). We derive interactions between virtual defects by decomposing the microscopic fields into their contributions from the each domain ($c = \sum c_i$, etc.), valid when the domains are well separated ($\vert \mathbf{R_i} - \mathbf{R_j}\vert > R_a$), and define the centre of mass and polarity of the each cluster ($\mathbf{R}_i$ and $\mathbf{p}_i$) as before:
\bea \label{eq:effR}
&& \mathbf{\dot R}_i(t) = v_{c} \mathbf{p}_i(t) - \sum_{j\neq i} \f{W}{c_t^{(i)}}\int d^2r \mathop{} c_i(\mathbf{r}) \grad \rho_j\l \mathbf{r} - \mathbf{R}_{j} \r \nn\\
&& \mathbf{\dot p}_i(t) = \sum_{j\neq i} \f{\zeta}{c_t^{(i)}}\int d^2r \mathop{} c_i(\mathbf{r}) \grad \rho_j(\mathbf{r} - \mathbf{R}_{j})
\eea
The second equation must be augmented with the constraint that the magnitude of each $\mathbf{p}_i$ is constant.

\indent We develop the integral appearing in both equations into a multipole expansion, valid when the length scale over which the myosin cloud varies, $r_0$, is larger than the characteristic size of the virtual defect. Expanding $\rho_j(\mathbf{r} - \mathbf{R}_{j})$ around $\mathbf{r} = \mathbf{R}_i$ we find:
\bea \label{eq:interaction}
\int d^2r \, c_i(\mathbf{r}) \, && \grad \rho_j(\mathbf{r} - \mathbf{R}_j) = M^{(0)} \grad \rho_j(\mathbf{R}_{ij}) \nn \\
&& + \l \mathbf{M}^{(2)}\l \mathbf{R}_i \r : \grad \grad \r \grad \rho_j(\mathbf{R}_{ij}) + \cdots
\eea
where the multipole moments are $M^{(0)} = \int d^2r \, c_i(\mathbf{r})$ and $M^{(2)}_{\alpha \beta} (\mathbf{r^{\prime}}) = \f{1}{2} \int d^2r \, \l r_{\alpha} - r^{\prime}_{\alpha} \r \l r_{\beta} - r^{\prime}_{\beta} \r \, c(\mathbf{r})$. Expanding the myosin field into axisymmetric circular harmonics $\rho(\mathbf{R}) = \rho^{(0)}(R) + \rho^{(1)}(R) \f{\mathbf{p} \cdot\mathbf{R}}{R} + \ldots$ and using the forms given in Eqs.\,\ref{eq:profile}, \ref{eq:axiprofile}, we may close the effective equations of motion for a population of interacting virtual defects. Retaining for simplicity only the radially symmetric term in $\rho$ (i.e., Eq.\,\ref{eq:profile} and the monopole term in Eq.\,\ref{eq:interaction}), and using the shorthand $\mathbf{R}_{ij} = \mathbf{R}_i - \mathbf{R}_j$:
\bea \label{eq:effRLowest}
&& \mathbf{\dot R}_i(t) = v_{c} \mathbf{p}_i(t) - \sum_{j\neq i} \f{W A_1}{r_0} \text{K}_1 \l \f{|\mathbf{R}_{ij}|}{r_0} \r \hat{\mathbf{R}}_{ji} \nn\\
&& \mathbf{\dot p}_i(t) = \sum_{j\neq i} \f{\zeta A_1}{r_0} \text{K}_1 \l \f{|\mathbf{R}_{ij}|}{r_0} \r \hat{\mathbf{R}}_{ji} \cdot \l \mathds{1} - \mathbf{p}_i\mathbf{p}_i \r
\eea
Note that, as each $\rho_j$ decays away from $R_j$, the signs of $W$ and $\zeta$ give rise to an attractive `force' (\textit{phoresis}) and an attractive `torque' (\textit{taxis}), respectively, between virtual defects \cite{SuropriyaColloid,YoshinagaCollision}. We note that while only a virtual defect may exhibit taxis (which requires a non-zero $\mathbf{p}$) the imposition of a gradient of myosin motors may elicit the phoresis of an otherwise static aster (Movie 7 \cite{suppinfo}).

\indent We now turn to the consequences of Eqs.\,\ref{eq:effR} and \ref{eq:interaction} on the `scattering' of virtual defects. First, we investigate this in the geometry depicted in 
Fig.\,\ref{fig:scatter}(b) by numerically integrating the full equations of motion, Eqs.\,\ref{eq:ceqn}-\ref{eq:neqn}. Figure\,\ref{fig:scatter}(a) shows that for large impact parameter $b$ or for large self-propulsion speed $\propto v_0$ the two clusters `elastically scatter' off each other and escape to the far field (Movie\,3 \cite{suppinfo}), whereas for low $b$ or low $v_0$ they merge 
(Movie\,4 \cite{suppinfo}). Interestingly, as seen in the trajectories shown in Fig.\,\ref{fig:scatter}(a), the deflection angle of scattering decreases with $v_0$ even for a fixed $b$. This is reminiscent of the scattering of inertial particles, even though the `extended particles' here are formed by constituents undergoing overdamped dynamics.

\indent The symmetry of the scattering geometry of Fig.\,\ref{fig:scatter}(b) can be exploited to develop an analytic representation of the scattering process. We define: the distance of one virtual defect from the other, $r_c = \vert \mathbf{R}_{21}\vert$, and the radial projection of the polarity vector, $\gamma = \mathbf{R}_{12}\cdot\mathbf{p}_1/r_c$. 
\bea \label{eq:dynamicalsystem}
&& {\dot r}_c \, = 2v_c\gamma + 2\overbar{W} \, \text{K}_1\l \f{r_c}{r_0} \r \nn \\
&& {\dot \gamma} \, = (1- \gamma^2) \l 2\f{v_c}{r_c} - \overbar{\zeta} \, \text{K}_1\l \f{r_c}{r_0} \r \r
\eea
where $\overbar{W} = WA_1/r_0$ and $\overbar{\zeta} = \zeta A_1 /r_0$.

\indent A representative phase portrait of this dynamical system is depicted in Fig.\,\ref{fig:scatter}(c). It can be shown that Eqs. (\ref{eq:dynamicalsystem}) admit a single saddle node \cite{Strogatz} which is always between $0 < \gamma \leq 1$. Thus, in the physically relevant region of the phase plane $-1 < \gamma < 1$, the unstable separatrix of this saddle node acts as a phase boundary between merger and elastic scattering. 

\indent Note, in particular, that all trajectories escape to infinity or end in a merging -  there exist no bound states.

\begin{figure}
\begin{centering}
\includegraphics[scale=.05]{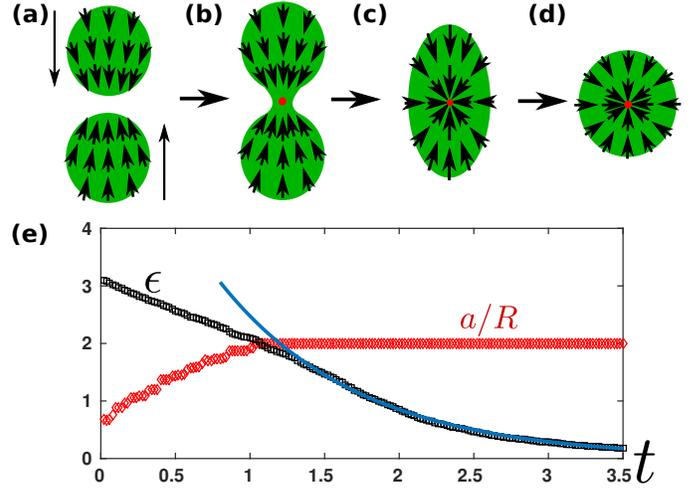}
\end{centering}
\caption{\label{fig:mergeschem} A schematic of the sequence of events in merger, namely \textbf{(a)} approach of the clusters, followed by \textbf{(b)} neck formation, which \textbf{(c)} grows to form an elliptic domain that \textbf{(d)} relaxes exponentially to a radially symmetric domain. \textbf{(e)} Time evolution of the ratio of the neck width $a$ to the single domain radius $R$ (red diamonds) as well as the shape factor $\epsilon = d_{\text{maj}}/d_{\text{min}} - 1$ (black squares), where $d_{\text{maj}}$, $d_{\text{min}}$ are the lengths of the major and minor axes of the ellipse enclosing the merging domain area. The blue line is an exponential fit.}
\end{figure}

\indent We end with a brief discussion on the sequence of events that occur when two domains merge. We find numerically that when two clusters merge in a head-on collision they do so in the following stereotypic manner (Fig.\,\ref{fig:mergeschem}(a) and Movie\,6 \cite{suppinfo}) - (i) initial overlap of the concentration fields leads to the formation of a neck region, which develops an aster defect (ii) the neck widens to produce an elliptic domain which then (iii) relaxes exponentially to a radially symmetric domain with an aster defect (as in \cite{HymanMerger}), Fig.\,\ref{fig:mergeschem}(b).

\indent When $K$ is large the aster core is unstable \cite{PetteyLubensky} and moves out of the merged domain. However, when $K$ is small, the defect core is stable and remains to form a stationary aster or spiral aster (contrast Movies\,4 and\,5 \cite{suppinfo}). In the intermediate regime, we find that the fate of the domain depends on the impact parameter $b$ and the self-advection speed $v_0$,
revealing the dynamical nature of the interacting myosin field. We will carry out a detailed study of the dynamics of merger in a later work.

\indent In summary, we have conducted a detailed study of the form and dynamics of the compact, emergent structures that arise in a hydrodynamic description of an active polar fluid with associated `motor' elements. By explicit construction of effective equations of motion, we have demonstrated that these structures behave as extended self-propelled particles that exhibit both phoresis and taxis.
We expect our results to be relevant to systems exhibiting both self-propulsion as well as contractile stresses, such as might be expected for the dynamics of actomyosin structures as the cell surface \cite{Gowrishankar}. We look forward to experimental tests of our predictions in \textit{in vivo} as well as \textit{in vitro} contexts \cite{Darius}.

\indent We thank S. Ramaswamy, K. VijayKumar, Amit Singh, 
B. R. Ujwal and the members of the Simons Centre for the Study of Living Machines for insightful
discussions.

\end{document}


\begin{center}
\Large{\bf Emergent Structures in an Active Polar Fluid : dynamics of shape, scattering and merger \\ \textit{Supplementary Information}}
\vskip 0.2in
\Large{Kabir Husain$^{1}$ and Madan Rao$^{1,2,\ast}$\\
\vskip 0.1in
\normalsize{$^{1}$Simons Centre for the Study of Living Machines, National Centre for Biological Sciences (TIFR), Bangalore 560 065, India}\\
\normalsize{$^{2}$Raman Research Institute, C.V. Raman Avenue, Bangalore 560080, India}\\
\vskip 0.1in
\normalsize{$^\ast$To whom correspondence should be addressed; E-mail:  madan@ncbs.res.in}
}
\end{center}

\tableofcontents

\section{Model}

\subsection{Equations of Motion}

\indent Here we construct the basic equations for the active hydrodynamics of polar actin filaments and myosin based on symmetries and conservation laws \cite{ActiveReview}, and then describe the approximations that lead finally to dynamical equations displayed in Eqs.\,(1) in the main text. To describe a population of dynamic actin filaments and associated bound myosin minifilaments embedded in a static cortical meshwork \cite{Gowrishankar}, we use the following hydrodynamic variables: the actin filament density $c(\mathbf{r},t)$, the actin filament orientation $\mathbf{n}(\mathbf{r},t)$, the bound myosin density $\rho(\mathbf{r},t)$, and the hydrodynamic velocity $\mathbf{v}(\mathbf{r},t)$.

Ignoring actin turnover, 
the density of the polar actin filaments is conserved.
\bea
\label{eq:activec1}
\p_t c = -\grad\cdot\mathbf{J}_a \nn \\
\mathbf{J}_a = -D_a(\rho)\grad c + c\mathbf{v} + v(\rho,c) c \mathbf{n}
\eea
where $D_a(\rho=0) > 0$ and $v(\rho=0)=0$, the latter being a purely active contribution \cite{Ahmadi}. This latter term, arising from the walking of myosin minifilaments that cross-link the dynamic actin filaments to each other and to the static meshwork, is then expanded to lowest order around the mean actin and myosin densities $c_0$, $\rho_0$:
$v(\rho,c) \approx v(\rho_0,c_0) \equiv v_0$.


\indent The dynamics of bound myosin motors is described in terms of its current $\mathbf{J}_m$ as well as turnover,
\bea
\label{eq:activem1}
\p_t \rho = -\grad\cdot\mathbf{J}_m + k_{\text{bind}}(c) - k_u\rho \nn \\
\mathbf{J_m} = -D_m(c,\rho)\grad \rho + v_m(\rho,c) \rho \mathbf{n} + \rho\mathbf{v}
\eea
where again we have taken $v_m(\rho,c) \approx v_1$, and have assumed an unlimited bath of unbound myosin.  The origin of the advective term $\propto \mathbf{n}$ in the myosin current arises from myosin action against a static substrate: the sign and magnitude of $v_1$ will depend on the microscopic details. In the limit of high friction against the substrate, $v_1$ must be either positive or negative with small magnitude - to reduce the number of free parameters, we take $v_0 = v_1$, but this is not necessary. The dependence of the binding rate of myosin on filament density is taken to be $k_{\text{bind}}(c) = k_b c/(c+c_h)$, which is linear for small $c$ and saturates for $c > c_h$.

\indent The equations for the hydrodynamic velocity $\mathbf{v}$ of the Stokesian fluid, dominated by friction of the cortex,
is given by force balance - 
 \bea
\label{eq:activev1}
\Gamma \mathbf{v} = \grad \cdot\sigma^a \nn \\
\sigma_{ij}^{a} = -W(c,\rho) \approx -W\rho + \mathcal{O}\l \rho^2,\, \rho c \r
\eea
where we have assumed that the total stress is dominated by the active contributions. 
$W < 0$ corresponds to a contractile stress. The velocity field can be approximately solved for from $\Gamma \mathbf{v} \approx \grad \cdot\sigma^{a}$, obtaining $\mathbf{v} \sim \nab \rho$, which we insert into the filament and motor currents, renormalising the diffusion co-efficient in the latter (which we do not consider further) and adding a advective term $\sim c \grad\rho$ to the former. The actin density $c$ enters into the myosin current through higher order contributions to the active stress, Eqs. (\ref{eq:activev1}), and so we do not consider those terms here.

\indent The filament orientation is subject to forces and torques arising from thermodynamic sources, hydrodynamic flows and activity and its dynamics are given by \cite{ActiveReview,muhuri},
\be \label{eq:activen}
\frac{D \mathbf{n}}{D t} +
\lambda(\rho) \mathbf{n}\cdot\grad \mathbf{n} + \ldots = K_1(\rho) \nabla^2\mathbf{n} + K_2(\rho) \grad({\grad}\cdot\mathbf{n}) - \zeta_0 \grad c + M(\rho) \grad \cdot \sigma^a + (\alpha(c) - \beta(c) n^2)\mathbf{n}
\ee
where on the left, $\frac{D \mathbf{n}}{D t} = \f{\p \mathbf{n}}{\p t} + \mathbf{v} \cdot \grad\mathbf{n} - \f{1}{2} (\grad \times \mathbf{v}) \times \mathbf{n}$ is the derivative in the comoving and co-rotating frame, and the $\ldots$ are the rest of the active convective nonlinear terms \cite{TonerTu} and the flow-alignment terms. 

\indent The first two terms on the right hand side correspond to restoring forces against bend and splay configurations with Frank coefficients $K_1$ and $K_2$, while the $\alpha^{\ast}(c)$ and $\beta^{\ast}(c)$ terms act as a soft ordering potential that favours a non-zero director when the density field $c$ is greater than the Onsager value $c_{\ast}$. As shown
in \cite{TonerTu},  $\zeta_0 > 0$ as a consequence of filament compressibility. $M(\rho)$ is a particle
mobility and is taken to be positive, reflecting the contractile nature of myosin induced active stresses, and constant: $M(\rho) \approx \zeta > 0$.



\indent The soft potential controlling the lyotropic ordering transition is defined by $\alpha^{\ast}(c) = \nu(c/c_{\ast} - 1)$ and $\beta^{\ast}(c) = \nu(1+c/c_{\ast})$, which ensures that $\vert {\bf n}\vert^2 \to 1$ when $c \gg c_{\ast}$. For simplicity, we work in the one constant approximation where $K_2 = 0$ \cite{degennes}. The final equations we study are,
\bea
&& \f{\p c}{\p t} = -\grad\cdot \l -D_a\grad c + v_0 c \mathbf{n} - W c \grad \rho \r \nn \\
&& \f{\p \rho}{\p t} = -\grad\cdot \l -D_m\grad \rho + v_0\rho\mathbf{n} \r + k_{\text{bind}}(c) - k_u \rho \nn \\
&& \frac{\p \mathbf{n}}{\partial t} = K \nabla^2\mathbf{n} - \zeta_0 \grad c + \zeta \grad \rho + (\alpha^{\ast}(c) - \beta^{\ast}(c) n^2)\mathbf{n}
\eea
and appears as Eq.\,1 in main text.

\subsection{Integrating Out Myosin}

\indent It is instructive to consider the limit in which our equations reduce to earlier descriptions of cortical actomyosin \cite{Gowrishankar,KripaPhase}. In those, the dynamics of myosin were not treated explicitly but instead entered the equations as effective interactions between the active actin filaments. Correspondingly, if we slave the bound myosin density to the filament density and orientation and solve for $\rho$ in the mean-field limit (i.e., determined entirely by binding-unbinding), we obtain $\rho \sim c$, to lowest order. Plugging this back into the equations for $c$ and $\mathbf{n}$, we find that the diffusive part of the filament current receives a higher order correction ($D_a \to D_a(c)$) and that the anchoring terms collapse into one, $-\zeta_0\grad c + \zeta\grad\rho \to \zeta_{\text{eff}}\grad c$. Neglecting the nonlinear correction to the filament diffusion, we get
\bea
&& \f{\p c}{\p t} = -\grad\cdot\l -D_a\grad c + v_0 c \mathbf{n} \r \\
&& \f{\p \mathbf{n}}{\p t} + \lambda \l \mathbf{n}\cdot\grad \r\mathbf{n} = K_1\nabla^2\mathbf{n} + K_2\grad\l \grad\cdot\mathbf{n} \r + \zeta_{\text{eff}}\grad c + (\alpha(c) - \beta(c)n^2)\mathbf{n}
\eea
If $\zeta$ is large compared to $\zeta_0$, the effective anchoring co-efficient $\zeta_{\text{eff}}$ is positive and we recover the earlier equations \cite{Gowrishankar,KripaPhase}. Using these 
equations, we had described the emergent behaviour of static, compact asters \cite{Gowrishankar,KripaPhase}, and hence we expect our system to produce the same in an 
appropriate limit (Fig.\,1(a) of main text).

\section{Linear Calculations}

\subsection{Stability Analysis of the Homogeneous Disordered Phase}

\indent The equations of motion admit homogeneous steady states in which $c(\mathbf{r},t) = c_0$, $\rho(\mathbf{r},t) = \rho_0$ and $\mathbf{n}(\mathbf{r},t) = \mathbf{n}_0$, where $n_0 = 0$ corresponds to the isotropic phase and non-zero $n_0$ corresponds to the polar ordered phase. We ask if the homogeneous isotropic phase is linearly stable with respect to perturbations: $c(\mathbf{r},t) = c_0 + \delta c(\mathbf{r},t)$, $\rho(\mathbf{r},t) = \rho_0 + \delta \rho(\mathbf{r},t)$ and $\mathbf{n}(\mathbf{r},t) = \delta \mathbf{n}(\mathbf{r},t)$, where $\rho_0 = k_{\text{bind}}(c_0)/k_u$ and $c_0 < c_{\ast}$.

\indent Inserting the perturbed fields into the equations of motion, linearising and using the wave ansatz $f(\mathbf{r},t) =  \fou{f}(t) \exp \l -\iu \mathbf{k}\cdot\mathbf{r} \r$,

\be
\p_t \begin{pmatrix}\delta \fou{c} \\ \delta \fou{n}_x \\ \delta \fou{n}_y \\\delta \fou{\rho}_b \end{pmatrix} = \mathbf{M_0} \begin{pmatrix}\delta \fou{c} \\ \delta \fou{n}_x \\ \delta \fou{n}_y \\ \delta \fou{\rho}_b \end{pmatrix}
\ee

\indent The growth rate of the perturbation is given by decomposing the matrix $M_0$ into its eigenvectors and examining the real part of the corresponding eigenvectors. The most general case is cumbersome to work with, so we consider the following limits. We set $D_m = D_a$ and consider the regime in which $k_{\text{bind}}(c) \approx k_b c$ for simplicity. We analyse first the potential instability driven by myosin's advection of actin filaments and set $\zeta$, $\zeta_0$ and $v_0 \to 0$. In that limit the four eigenvalues $\omega_i$ of $\mathbf{M_0}$ are given by:

\bea
&& \omega_{1,2} = -K k^2 + \alpha \\
&& \omega_{3,4} = \f{1}{2}\l -k_u - 2D_a k^2 \pm \sqrt{k_u^2 + 4c_0k_b \vert W\vert k^2} \r
\eea

\indent The first two eigenvalues describe the relaxation of the orientation field $\mathbf{n}$ to $0$. The second two describe the coupled dynamics of $\delta c$ and $\delta\rho$. Inspection of the eigenvalues reveals that an instability occurs when:

\be
\f{|W| c_0 k_b}{D_a k_u} > 1
\ee

\indent A band of wavevectors between $k = 0$ and $k = \sqrt{\vert W\vert k_bc_0 - D_a k_u}/D_a$ becomes unstable. The fastest growing wavevector is given by:

\be
k_{\text{max}} = \f{\sqrt{(Wk_bc_0)^2 - D_a^2k_u^2}}{2D_a\sqrt{\vert W\vert k_bc_0}}
\ee

\indent Numerical solution of the equations of motion in this limit reveals that the instability grows to form static clumps of actin and myosin, which may either be disordered ($\langle \mathbf{n}\rangle$ remains $0$ everywhere) or ordered (see section \ref{localord}).

\indent Alternatively, the system may be studied in the limit $W$, $\zeta_0 \to 0$. In this case, the eigenvalues of $\mathbf{M_0}$ are found to be:

\bea
&& \omega_1 = -K k^2 + \alpha \\
&& \omega_2 = -D_a k^2 + k_u \\
&& \omega_{3/4} = \f{1}{2} \l \alpha - (D_a + K)k^2 \pm \f{1}{k_u} \sqrt{k_u^2\l \alpha + (D_a - K)k^2 \r + 4c_0kbk_uv_0\zeta k^2} \r 
\eea

\indent The criterion for instability is found to be $v_0\zeta k_b c_0 > D_a |\alpha| k_u$. Once again, a band of wavevectors between $k=0$ and $k = \sqrt{\f{D_a k_u \alpha + c_0 k_b v_0 \zeta}{D_a K k_u}}$ is unstable. The fastest growing wavevector is given by:

\be
k_{\text{max}}^2 = \f{-D^-D^+ k_u \alpha - 2D_a K \overbar{\zeta} + D^+\sqrt{\overbar{\zeta}D_a K (D^-k_u\alpha + \overbar{\zeta})}}{D_a (D^-)^2 K k_u}
\ee

\noindent where $\overbar{\zeta} = k_b c_0 v_0 \zeta$ and $D^{\pm} = D_a \pm K$.

\indent This instability involves a non-zero splay component of $\mathbf{n}$ (i.e., $\grad \cdot \mathbf{n}\neq 0$), due to the coupling of the perturbations of the density fields $\delta c$, $\delta \rho$ to the longitudinal component of orientation, $\mathbf{k}\cdot\delta \mathbf{n}$. Correspondingly, the results above can also be obtained by solving the linear system for the variables $\delta c$, $\delta \rho$ and $\mathbf{k}\cdot\delta \mathbf{n}$. This indicates that the resultant pattern may be clumps of actin and myosin with an aster texture, which is confirmed by numerical integration.

\subsection{Local Ordering} \label{localord}

\indent Even in the absence of the anchoring terms, filaments within clusters may order at sufficiently high myosin contraction, even though the nucleation of the platform occurs without breaking local orientational symmetry and $\mathbf{n}$ is $0$ in the early stages of cluster formation. As the local density of filaments rises it may cross the Onsager value $c_{\ast}$ and the filaments inside the structure may be susceptible to spontaneous orientational ordering. Assuming an ansatz $c(\mathbf{r}) = c_0 > c_{\ast}$ for $r < R_0$ and $c_b < c_{\ast}$ otherwise, we perturb the homogeneous state $\mathbf{n} = 0$ within the cluster (i.e., $r < R_0$). In the absence of anchoring, the orientation of the isotropic state is unstable to perturbations of wavelength larger than $1/k_n$, where $k_n^2 < \alpha(c_0-c_{\ast})/K$. The disordered state of filaments within a cluster is then unstable for clusters of radius larger than $R_c$, where $R_c = \sqrt{\f{K}{\alpha^{\ast}(c_0)}}$. Upon ordering, the Frank term ensures that the filaments within a cluster are all oriented in a single direction.

\section{Numerical Calculations}

\subsection{Parameter Values}

\indent The equations were made dimensionless by scaling with chosen characteristic units of space and time:

\indent \textbf{Time:} Unit of time $t = \nu^{-1}$, with $\nu \sim 1$ s$^{-1}$, from an estimate of the rotational diffusion time of an actin filament of length $\sim .1 \to 1 \, \mu m$ \cite{marchettispiral}.

\indent \textbf{Length:} Units of length $l = \sqrt{D_a/\nu}$. From previous FCS studies in living cells, the diffusion coefficient of dynamic actin filaments in the cortex may be estimated to be $D_a \sim 0.1 \, \mu $m$^2$ s$^{-1}$ \cite{Gowrishankar}. Hence $l \sim 0.3 \, \mu$m.

\begin{center}
\begin{tabular}{ |c|c|c|c| } 
 \hline
 Parameter (Dimensions) & Scaled Value & Physical Value \\ 
 \hline
 $v_0$ ($l/t$) & $0.1 \to 2.0$ & $ 0.03 \to .6 \, \mu$m s$^{-1}$ \\
 $\zeta$ ($l^3/t$) & $0 \to 30$ & $ 0 \to 1 \, \mu$m$^3$ s$^{-1}$ \\
 $K$ ($l^2/t$) & $0.2 \to 2.0$ & $ 0.02 \to .2 \, \mu$m$^2$ s$^{-1}$ \\
 $W$ ($l^4/t$) & $-10 \to -30$ & $ 0.1 \to .3 \, \mu$m$^4$ s$^{-1}$ \\
 $k_b$ ($l^{-2}/t$) & $0.25 \to 16$ & $ 2.5 \to 160 \, \mu$m$^{-2}$ s$^{-1}$ \\
 $k_u$ ($1/t$) & $0.125 \to 8$ & $ 0.125 \to 8 \,$ s$^{-1}$ \\
 \hline
\end{tabular}
\end{center}

\indent These values are consistent with those used in earlier studies \cite{Gowrishankar,Pragya}, and have been chosen so that all terms are of similar order and are in the appropriate pattern-forming regime as dictated by the linear stability analysis.

\subsection{Single Cluster Phase Diagram} \label{sec:singlephase}

\indent The equations of motion were solved on a square lattice with an explicit Forward-Time-Central-Space (FTCS) scheme, with a grid size of $10\times 10$ diffusion lengths and spacing $dx = 1/8$. All numerical calculations were performed with peridic boundary conditions. A single cluster was nucleating by using a gaussian initial condition for $c$ with initial amplitude $c_0 = .1$ and width $l = 4$, with a small amount of initial noise added. The remaining parameters were fixed at the following: $D_m = D_a$, $\alpha = 1$, $c_{\ast} = 1/64$, $v_0 = 0.1$, $W = -30$, $k_b = 1/4$, $k_u = 1/8$, $c_h = 1/4$.

\indent The following quantities were used to distinguish between the phases and construct the phase diagram. The angular brackets indicate an average over space.

\begin{itemize}
\item $|\langle c \mathbf{n} \rangle| = 0$, $|\langle c \grad\times\mathbf{n} \rangle| = 0$ and $\langle c \grad\cdot\mathbf{n} \rangle < 0$ corresponding to static asters.
\item $|\langle c \mathbf{n} \rangle| = 0$, $|\langle c \grad\times\mathbf{n} \rangle| = 0$ and $\langle c \grad\cdot\mathbf{n} \rangle > 0$ corresponding to static anti-asters.
\item $|\langle c \mathbf{n} \rangle| = 0$ and $|\langle c \grad\times\mathbf{n} \rangle| > 0$ corresponding to static spiral asters.
\item $|\langle c \mathbf{n} \rangle| > 0$ and $\langle c \grad\cdot\mathbf{n} \rangle > 0$ corresponding to motile virtual anti-asters.
\item $|\langle c \mathbf{n} \rangle| > 0$ and $\langle c \grad\cdot\mathbf{n} \rangle < 0$ corresponding to motile virtual asters.
\end{itemize}

\indent The phases with their different textures are displayed in Fig.\,\ref{fig:textures}.


\subsection{Two-Body Scattering}

\indent The parameters used in the numerical calculations for two body scattering (as shown in the main text, Figure 2) were: $D_m = 1$, $D_a = 2$, $\zeta_0 = 1$, $\nu = 1$, $c_{\ast} = 1/64$, $W = -7$, $\zeta = 7$, $K = 2$, $k_b = 16$, $k_u = 8$, $c_h = 1/4$.

\indent We also carried out two-body scattering with the modified parameters $W = -15$, $\zeta = 5$, $k_b = 1$, $k_u = 0.5$ and $K = 3$, presented in Fig. \ref{fig:second}(c), sampling impact parameter values from $b = 0$ to $b = 15$ in steps of $0.5$ and $v_0$ values from $0.4$ to $2.0$ in steps of $0.1$.

\subsection{Merging}

\indent The parameters used in the numerical calculations for merging were: $D_m = 2$, $D_a = 1.85$, $\zeta_0 = 1$, $\nu = 1$, $c_{\ast} = 1/64$, $W = -6$ to $-10$, $\zeta = 7$, $K = 3$, $k_b = 4$, $k_u = 2$, $c_h = 1/4$.

\begin{figure}[h]
\begin{centering}
\includegraphics[scale=.2]{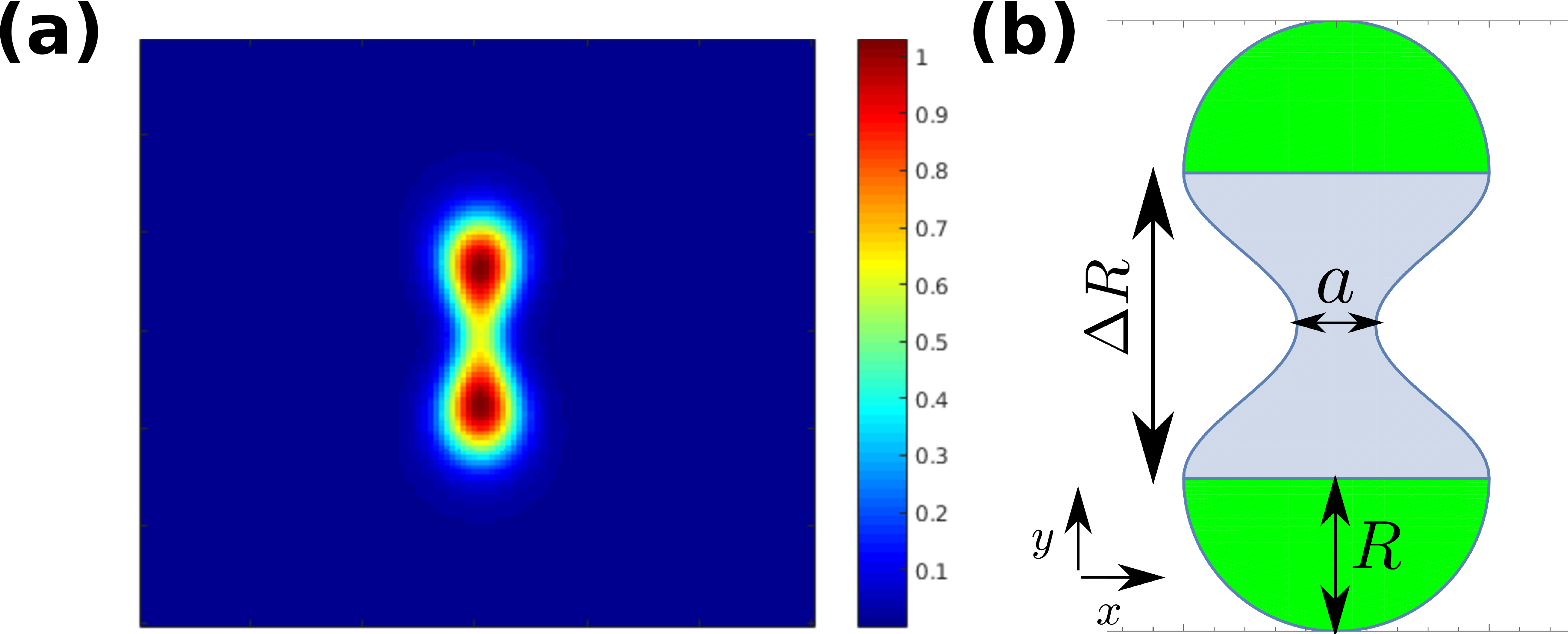}
\caption{\label{fig:mergeansatz} \textbf{(a)} Snapshot of the filament density field $c$ during domain merger. \textbf{(b)} Schematic of the form used to fit the domain area - see text for details.}
\end{centering}
\end{figure}

\begin{figure}[h]
\begin{centering}
\includegraphics[scale=.1]{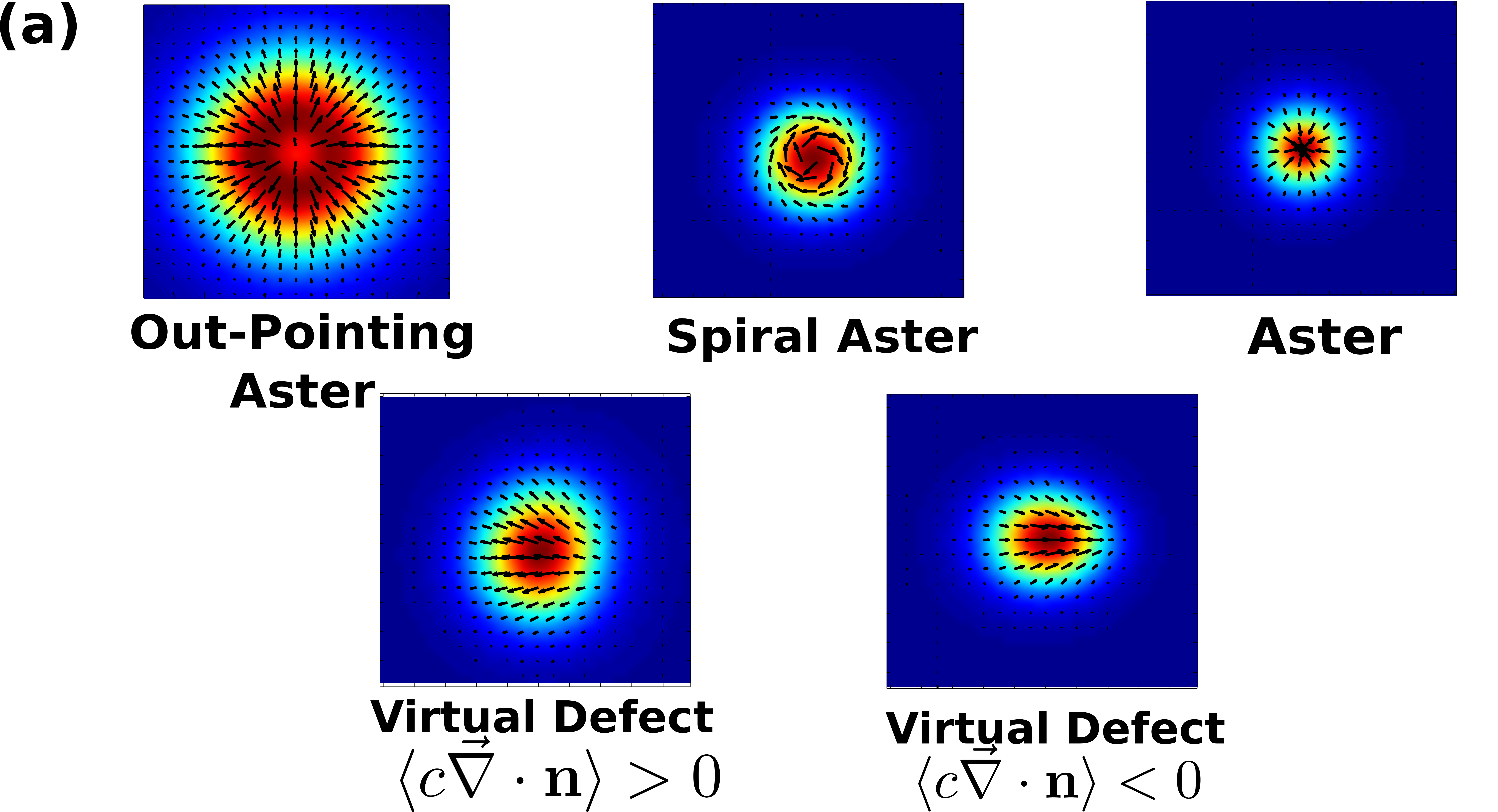}
\caption{\label{fig:textures} \textbf{(a)} Same as in Fig.\,1(a) in the main text: finite-size textures found in numerical calculations.}
\end{centering}
\end{figure}

\begin{figure}[h]
\begin{centering}
\includegraphics[scale=.14]{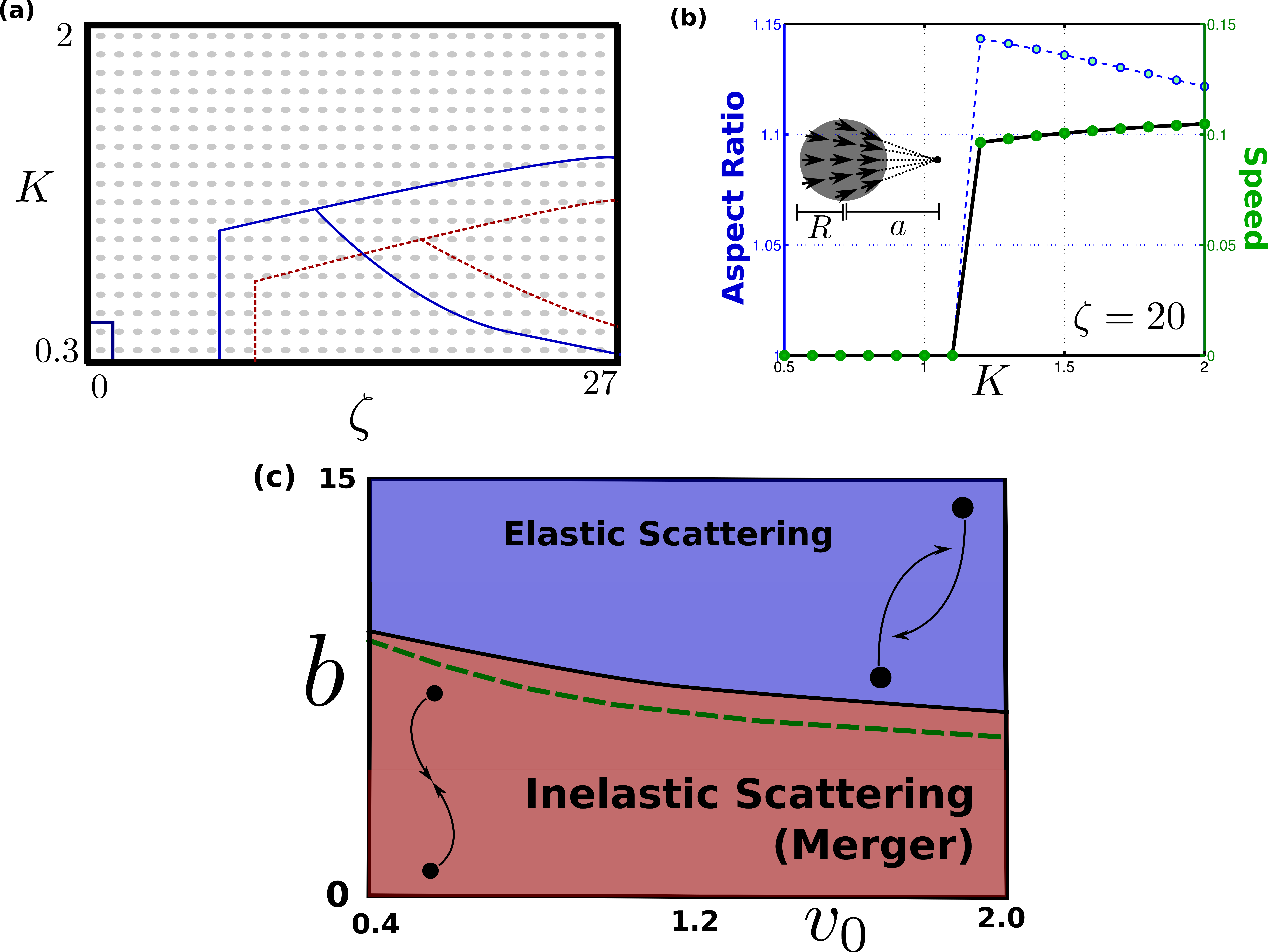}
\caption{\label{fig:second} \textbf{(a)} Phase boundaries with $\lambda = 0$ (blue lines; same as in Fig.\,1(a) in main text) and $\lambda = v_0 = 0.1$ (red dashed lines). \textbf{(a)} Speed and aspect ratio as a function of $K$ (with $\zeta = 20$ and $v_0 = 0.1$). Inset: Schematic of the \textit{ansatz} used to represent a virtual defect - arrows represent $\mathbf{n}$, the dot represents the core of the virtual defect, and the grey circular domain represents the filament and myosin concentration (assumed to be uniform inside the domain). \textbf{(c)} Outcome of two-body scattering as a function of impact factor $b$ and self-advection $v_0$ (see Figure 2 in main text). Red and blue regions show regimes of merger and scattering, respectively, determined from full numerical integration of the hydrodynamic equations. Dashed green line shows boundary as predicted by the effective theory, i.e. Eq. (10) in the main text.}
\end{centering}
\end{figure}

\indent To analyse the shape deformations during the merging of two domains we threshold the filament density $c$ and fit it to the parameterised area depicted in Fig.\,\ref{fig:mergeansatz}(b). This \textit{ansatz} consists of two semicircular regions of radius $R$ separated by a distance $\Delta R$. The intervening region satisfies the inequality,
\be
\vert x \vert < R\l 1 - \f{1-\f{a}{2R}}{2} \l 1 + \cos\l\f{2 \pi y} {\Delta R} \r \r \r \nn
\ee

\noindent where $x$ and $y$ are co-ordinates defined with the origin lying at the geometric centre of the area. This form ensures that the boundary of the domain has a continuous first derivative everywhere.

\section{Movie Captions}

\begin{enumerate}

\item \textbf{Movie 1: } Movie shows the time evolution of an initially homogeneous, orientationally isotropic phase into a population of asters. Here, the colours represent the local filament density $c$ and the vectors represent the magnitude and direction of the filament orientation $\mathbf{n}$. Note that individual asters are static and move only when under the influence of others - i.e., when merging. The simulation size is $30\times30$ in units of $\sqrt{D_a/\alpha}$. Parameters used were: $D_m = D_a = 1$, $K = 0.1$, $\zeta_0 = 1$, $\zeta = 25$, $c_{\ast} = 1/64$, $v_0 = 0.1$, $W = 30$, $k_b = .25$, $k_u = .125$, $c_h = 1/4$.

\item \textbf{Movie 2: } Movie shows the evolution of an initially homogeneous, orientationally isotropic phase into an evolving population of interacting, merging virtual defects. Once again, the heatmap represents $c$ and the vectors $\mathbf{n}$. Note, in contrast with movie 1, that the virtual defects are self-propelled and move in the direction of their own net polarity. To show many structures, here the size of the simulation is larger than movie 1: $140\times 140$ in units of $\sqrt{D_a/\alpha}$. Parameters used were: $v_0 = 0.2$, $D_m=1.0$, $c_{\ast} = 0.5$, $\zeta_0 = 0$, $\zeta = 5$, $K = 1.5$, $W = 25$, $k_b = 0.25$, $k_u = 0.125$, $c_h=0.25$.

\item \textbf{Movie 3: } Movie shows elastic scattering of a two virtual defects. Here, the heatmap represents $c$ and the vectors $\mathbf{n}$. Domains are nucleated with opposite polarities and varying impact parameter as described in the main text. Note here how the two structures not only attract each other, but also rotate to attempt to face each other - an example of both phoresis and taxis. Parameters as in section 3.2.

\item \textbf{Movie 4: } Movie shows inelastic merger of two virtual defects, with the merged domain becoming a virtual defect. The heatmap represents $c$ and the vectors $\mathbf{n}$. Note that the direction in which the merged domain moves is perpendicular to the axis along which the merger occurs. Parameters as in section 3.2.

\item \textbf{Movie 5: } Movie shows inelastic merger. Here, the heatmap represents $c$ and the vectors $\mathbf{n}$. In contrast with movie 4, here the merged domain becomes a static aster.

\item \textbf{Movie 6: } Movie shows inelastic merger as in Movie 4, but slowed down to highlight sequence of events in merger. In particular, note neck formation early in the merger process, and relaxation from an elliptical domain to a circular domain after the neck has widened. Note that, during this period, there exists an aster defect at the centre of the merging domain; in this example, the defect core moves out of the domain once merger has been completed to give rise to a virtual defect texture.

\item \textbf{Movie 7: } Movie shows the phoresis of an aster defect in an applied, static gradient of myosin ($\rho_s$, entering the equations of motion as $-Wc\grad \rho_s$ and $\zeta \grad \rho_s$ in the filament current and orientation equation, respectively). Parameters were as in Section \ref{sec:singlephase}, except for $W = -20$ and $v_0 = 0.2$, with $K = 1.0$ and $\zeta = 30$. The applied myosin gradient has magnitude $|\grad \rho_s| = 0.01$ in simulation units, and increases from top left to bottom right.

\end{enumerate}

